\documentstyle[preprint,prl,aps]{revtex}

\begin{document}
\draft

\title{
Hole and Pair Structures in the $t$-$J$ model
}
\author{ Steven R.\ White$^1$ and D.J.\ Scalapino$^2$}
\address{ 
$^1$Department of Physics and Astronomy,
University of California,
Irvine, CA 92717
}
\address{ 
$^2$Department of Physics,
University of California,
Santa Barbara, CA 93106
}
\date{\today}
\maketitle
\begin{abstract}
Using numerical results from density matrix renormalization group
(DMRG) calculations for the $t$-$J$ model, on systems as large as
$10\times7$, we examine the structure of the one and two hole ground
states in ladder systems and in two dimensional clusters.  A simple
theoretical framework is used to explain why holes bind in pairs in
two-dimensional antiferromagnets.
For the case $J/t=0.5$, which we have studied, the hole pairs reside
predominantly on a $2\times2$ core plaquette with the probability
that the holes are on diagonal sites greater than nearest-neighbor
sites.  There is a strong singlet bond connecting the spins on the two
remaining sites of the plaquette.
We find that a general characteristic of dynamic holes in an
antiferromagnet is the presence of frustrating antiferromagnetic bonds
connecting next-nearest-neighbor sites across the holes. Pairs of
holes bind in order to share the frustrating bonds.

At low doping, in addition to hole pairs, there are two additional
low-energy structures which 
spontaneously form on certain finite systems. The
first is an undoped $L\times2$ spin-liquid region, or ladder. The
second is a hole moving along a one dimensional chain of sites. 
At higher doping we expect that hole pairing is always favored.
\end{abstract}

\pacs{PACS Numbers: 74.20.Mn, 71.10.Fd, 71.10.Pm}

\narrowtext

\section{Introduction}
The main obstacle to understanding two-dimensional doped
antiferromagnets has been the inadequacy of current analytical
and numerical tools when applied to these systems. A number
of analytical approaches are available which work well in
either high dimensions (such as dynamical mean field
theory\cite{infinited}) or one 
dimension (such as bosonization\cite{bosonization}), but not in two dimensions.
Numerical approaches such as quantum Monte Carlo\cite{qmc} and exact 
diagonalization\cite{pre,hp,dag,prelov,poilblanc} have been very useful,
but quantum Monte Carlo suffers from a sign problem at low
temperatures\cite{sign},
and exact diagonalization can only be applied to small clusters.

Recently, however, density matrix renormalization group (DMRG)
techniques have been developed which allow one to obtain accurate,
detailed information about ground state expectation values on
significantly larger clusters\cite{dmrg,cavo}. 
We have performed DMRG calculations on a variety of
$t$-$J$ clusters. We have been able to treat systems of width 3 and 4
at a variety of dopings, with lengths up to 32 sites. At low doping,
we have results from wider systems, such as $10\times 7$. 
Here we examine the structure of the ground state of $t$-$J$
clusters doped with one or two holes.
Specifically, we have calculated the
ground state expectation value of the spin $S^z_i$ and the exchange
field $\vec S_i \cdot \vec S_j$ around a dynamic hole or pair of
holes. We have also calculated the spatial kinetic energy
distribution of one or two holes in a cluster, the spatial
kinetic energy distribution of one hole when a second hole has
been projected onto a particular site, and the hole-hole correlation
function in a two-hole state. From these calculations one obtains
new insight into the nature of the structures holes can induce in an
antiferromagnetic host and the origin of pair binding seen in some
clusters.  In a
subsequent paper we will discuss the effect of additional holes and
present results for pairing correlations
on long two, three, and four chain ladders.

\section{Theoretical Framework}
We first describe a simple theoretical framework for
understanding hole motion in a $t$-$J$ model.
The Hamiltonian is\cite{zhang}
\begin{equation}
{\cal H}= {\cal H}_S + {\cal H}_K = J \sum_{\langle ij\rangle} (\vec S_{i} \cdot
\vec S_{j}
- \textstyle{1\over4} n_{i} n_{j} )
- t \displaystyle \sum_{\langle ij\rangle,s} 
P_G(c^\dagger_{i,s}c_{j,s} + c^\dagger_{j,s}c_{i,s})P_G 
\label{hamiltonian}
\end{equation}
\noindent
where $\langle ij\rangle$ denotes nearest-neighbor sites, $s$ is a spin index,
$\vec S_{i}$ and $c^\dagger_{i,s}$ are
electron spin and creation operators, 
$n_i = c^\dagger_{i,\uparrow} c_{i,\uparrow} +
c^\dagger_{i,\downarrow} c_{i,\downarrow}$,
and the Gutzwiller
projector $P_G$ excludes configurations with doubly occupied sites.
In the calculations shown here, we set the hopping $t=1$ and the
exchange $J=0.5$.

Let $|\psi\rangle$ be the ground state of a particular $t$-$J$
system with $N$ sites and $N-m$ electrons. 
Define a hole projection operator for site $i$ as
$P_h(i) = {c}_{i,\downarrow} {c}^{\dagger}_{i,\downarrow} 
{c}_{i,\uparrow} {c}^{\dagger}_{i,\uparrow}$. $P_h(i)$ projects out
the part of a wavefunction in which site $i$ is vacant.
Although we call this vacant site a ``hole'', there is not
necessarily any spin associated with the vacancy:
in the one dimensional $t$-$J$ model, for example,
there is not. A better term might be ``dynamic vacancy'', but
the use of the term ``hole'' has now become fairly standard.
In some systems, such as even-leg ladders,
an extra spin-$1/2$ is bound
to the vacant site, forming a composite object with charge and spin,
which is sometimes called a ``quasiparticle''.
We define an operator $P_h(h)$, which projects out a particular
configuration of $m$ holes, as 
$P_h(h) = P_h(h_1)\ldots P_h(h_m)$, where
$h = (h_1, \ldots ,h_m)$, and $h_1< \ldots < h_m$.
We can then separate $|\psi\rangle$ into parts with specified hole
locations as
\begin{equation}
|\psi\rangle = \sum_h P_h(h) |\psi\rangle = \sum_h a_h |\psi_h\rangle,
\end{equation}
where $|\psi_h\rangle$ is a normalized wavefunction
with holes at the specified sites, and $a_h > 0$. 
The ground state energy is given by
\begin{equation}
E = \sum_h a_h^2 
\langle \psi_h | {\cal H}_S |\psi_h\rangle
+
\sum_h\sum_{h'} a_h a_{h'} \langle \psi_h | {\cal H}_K
|\psi_{h'}\rangle .
\label{energy}
\end{equation}
The first term we refer to as the exchange energy, denoted by $E_S$.
The second term in Eq. (\ref{energy}), the hopping energy or kinetic
energy, can be written as
\begin{equation}
E_K = - t \displaystyle \sum_{\langle ij\rangle,s} \sum_h a_h a_{h'}
 \langle \psi_h |c^{\dagger}_{i,s}c_{j,s} |\psi_{h'}\rangle ,
\label{ke}
\end{equation}
where the hole configurations $h$ and $h'$ are the same, except that
$h$ has a hole at site $j$ and $h'$ has one at site $i$.
In general, we consider two hole configurations $h$ and $h'$ 
{\it adjacent } if they differ by a near-neighbor hop of a
single hole.
Define the hopping overlap between $h$ and $h'$ as
\begin{equation}
O_{h,h'} = \displaystyle 
 \langle \psi_h |\sum_{\langle ij\rangle,s} (c^{\dagger}_{i,s}c_{j,s} +  
 c^{\dagger}_{j,s}c_{i,s}) |\psi_{h'}\rangle .
\label{oh}
\end{equation}
Clearly a necessary condition for $O_{h,h'}$ to be nonzero is that
$h$ and $h'$ are adjacent, in which case only one pair of sites
$i$, $j$ appears in the sum. If $h$ and $h'$ differ only
in the position of hole $m$, $h_m \ne h_m'$, then
\begin{equation}
O_{h,h'} = \langle \psi_h |\sum_s c^\dagger_{h_m',s}c_{h_m,s} ) 
|\psi_{h'}\rangle .
\label{ohtwo}
\end{equation}
It is easy to see that $|O_{h,h'}| \le 1$.
The kinetic energy can be written as
\begin{equation}
E_K = - t \displaystyle \sum_{h,h'} a_h a_{h'} O_{h,h'}
\label{oke}
\end{equation}

We see that we can view the ground state as the result of a
set of coupled variational calculations, where the exchange energy of
each wavefunction $|\psi_{h}\rangle$ is minimized, subject to
having as much overlap as possible with adjacent hole
configurations. For $t > J$, the interplay between the kinetic and 
exchange terms is interesting. 
In the low doping regime, since there are more exchange terms which come 
into play, the bulk spin behavior is dominated by exchange.
Close to any holes, however, since $t > J$, substantial
modifications of the local spin arrangements can occur.
At higher doping, the bulk spin behavior can be changed
substantially as well.

Using DMRG, we can study $|\psi_{h}\rangle$ directly: we calculate
$|\psi\rangle$, and then measure operators of the form
$A P_h(h)$ (or $P_h(h) A P_h(h)$), normalizing by 
$\langle \psi | P_h(h)|\psi\rangle$.  It is useful to use 
$A = \vec S_i \cdot \vec S_j$, where $i$ and $j$ are near a hole or pair
of holes. This measurement gives us a ``snapshot'' of the spin 
configuration around a dynamic hole.
If this expectation value is close to $-0.75$ for two sites $i$ and
$j$, we say that there is a ``singlet bond''
connecting $i$ and $j$, even if there is no term in the Hamiltonian
directly coupling $i$ and $j$. We use the terms ``antiferromagnetic
bond'', ``valence bond'', or just ``bond'' simply to indicate that
$\langle \vec S_i \cdot \vec S_j \rangle < 0$.
Of course, N\'eel order makes weak
``bonds'' connecting widely separated sites on opposite sublattices,
but we will be particularly concerned here with bonds connecting
nearby sites on the {\it same}  sublattice.

We can also take a
snapshot of the spin configuration using $A = S^z_i$, for a single
hole on an even number of sites. In that case, the ground state is
degenerate with $S^z=\pm 1/2$, so that the expectation value of 
$S^z_i$ in one of the ground states is finite.
One can also project out some of the holes, and use
$A = n_{i,s} = c^\dagger_{i,s}c_{i,s}$, to find out where the
unprojected holes are, or
$A = K_{ij} = -t\sum_s (c^\dagger_{i,s}c_{j,s} + c^\dagger_{j,s}c_{i,s})$,
to study their motion.

\section{Results}

The results in this paper were all calculated using the finite
system version of DMRG\cite{dmrg}, keeping track of transformation
matrices to construct the initial guess for each superblock 
diagonalization\cite{cavo}. From 200 to 800 states were kept per
block, with 800 states necessary for the $10\times7$ system.
We performed hundreds or thousands of measurements for each
system. Ordinarily storing all the operators needed to 
measure so many quantities would greatly 
increase the memory used by the program, but since the transformation
matrices contain a complete description of the approximate wavefunction
produced by the DMRG algorithm, the measurements could be performed at 
the end of the calculation, in manageable groups of 50 to 100.
The large number of measurements, at worst, doubled or tripled the
computation time.

\subsection{Single chain}
As a warmup exercise, consider the 1D $t$-$J$ model, with one hole.
One might consider as a variational ansatz for $|\psi_{h}\rangle$ 
a N\'eel arrangement of the electron spins, with one electron removed.
In this ansatz we have made a ``quasiparticle'', since an extra spin
$1/2$ is associated with the hole. However, this is a very
poor ansatz: $|\psi_{h}\rangle$ has no overlap with $|\psi_{h\pm 1}\rangle$.
Alternatively, one can arrange the spins as shown in Fig. 1(a),
with shifted N\'eel arrangements separated by the hole\cite{ogata}. 
There are two spin wavefunctions $|\psi_{h}\rangle$, plus translations: 
for $h$ odd (even), an up spin 
is to the left of the hole, while for $h$ even (odd), a down spin
is to the left of the hole.  
In this case
there is complete overlap, and the kinetic energy associated with 
the hole takes on the maximal (in magnitude) value $-2t$. 
This is a simple 
intuitive argument for spin--charge separation in a 1D $t$-$J$ model.
Since a single hole moves freely, it also
suggests that there is no kinematic reason for the binding of pairs
of holes, although for unphysically large $J/t$ the diagonal term in
Eq.(\ref{energy}) can cause binding.

A justification for considering these N\'eel configurations for the 
1D $t$-$J$ model is the existence of power-law decaying
antiferromagnetic correlations in the 1D Heisenberg model. Bond-bond
correlations $\langle \vec S_i \cdot \vec S_{i+1} \, \vec S_j \cdot
\vec S_{j+1} \rangle$ also decay as a power law, suggesting a valence bond
configuration as a complementary ansatz: valence bonds occupy odd (even) 
links to the left of the hole, and even (odd) links to the right,
as shown in Fig. 1(b). If one takes this configuration, and applies
$\sum_s c^{\dagger}_{i,s}c_{j,s}$ to move the hole to a neighboring
site, one obtains the configuration in Fig. 1(c), with a valence
bond straddling the hole. Consequently, if we let the valence bond
configuration of Fig. 1(b) define $|\psi_{h}\rangle$ for all odd
sites $h$, and let the configuration of Fig. 1(c) define
$|\psi_{h}\rangle$ for all even sites, then the hole moves
freely, with the kinetic energy taking on its maximal value $-2t$.

In Fig. 1(d)-(e), we show DMRG results for the bond strength
$A = \vec S_i \cdot \vec S_j$ for a single hole in a 15 site 1D chain,
with open boundary condtions. The width of the line corresponding to
each bond has been made proportional to the bond strength, as
indicated by the scale in the box. The maximum possible bond
strength is $-3/4$.  The boundaries induce dimerization
in the system, and the results are quite similar to the valence
bond configurations shown in Fig. 1(b)-(c).
It is also possible to obtain results which look like Fig. 1(a).
In Fig. 1(f), we show results for the $S^z=1/2$ ground state
of a system with an even number of sites and one hole. 
The excess spin $1/2$ is spread out over the lattice.

Particularly interesting is the strength of the bond across the 
hole in Fig. 1(e). 
In order to maximize the hopping overlap with adjacent hole configurations,
in addition to having antiferromagnetic correlations on
nearest-neighbor links, we
expect such correlations between {\it next-nearest-neighbor } 
sites $i$ and $j$ if there is a hole at site $k$ which is a 
nearest-neighbor to both $i$ and $j$. Such a valence bond 
becomes a nearest-neighbor link after one hop of
the hole to either site, since moving the hole also moves the
bond. For example, suppose the hole configuration $h$ has a hole at
site $k$, with $i$ and $j$ nearest-neighbor sites to $k$.
Let $h'$ be the hole configuration after the hole hops from $k$ to
$i$. Since $j$ and $k$ are nearest-neighbor sites, we expect
a strong antiferromagnetic bond between these sites in
$|\psi_{h'}\rangle$. In order to maximize the hopping overlap 
$O_{h,h'}$, there will also be a strong antiferromagnetic bond between sites
$i$ and $j$ in $|\psi_{h}\rangle$. This tendency applies to
two dimensions as well as one, and appears as an essential ingredient
for pair binding in ladders and two dimensions.

\subsection{Two chains}
We now consider a two chain ladder system, with identical couplings
along the legs and rungs, $t=1$ and $J=0.5$. We consider first
a single hole. In Fig. 2(a) we show bond strengths in
$|\psi_{h}\rangle$ in the vicinity of the dynamic hole for a $32 \times 2$
lattice, with $h$ on site $(16,1)$. As we argued above, 
the system has a tendency to form antiferromagnetic correlations
on next-nearest-neighbor sites around the hole. Except in one dimension,
this tendency introduces {\it frustration}, since sites on the same
sublattice tend to be parallel. A single Heisenberg spin chain 
becomes dimerized for sufficiently large frustrating next-nearest
neighbor interaction ($J' > 0.24J$). Similar dimerization is clearly
evident in the two bonds above the hole Fig. 2(a). 
The dimerization weakens one of the nearest-neighbor
bonds around the hole sufficiently to allow two of the 
next-nearest-neighbor links to form antiferromagnetic correlations.
Since the hole is not quite at the center of the system,
the figure need not be symmetric.

A single Heisenberg spin chain with frustrating next-nearest
neighbor interaction $J' > 0.5 J$ develops incommensurate, spiral
spin correlations, in addition to
dimerization\cite{bursill,chitra,zigzag}. We have looked for
this in $|\psi_{h}\rangle$ by measuring $(\vec S_i \times \vec S_j)
\cdot (\vec S_k \times \vec S_l)$, where the hole was on site
(16,1) and the spin operators $i$, $j$, $k$, and $l$ were for sites
(15,1), (15,2), (17,2), and (17,1), respectively. No enhancement of
this quantity by the presence of the hole was found, nor was any
found in the other lattices we studied here. However, it still might
occur\cite{siggia} in other parameter regimes, such as $t >> J$.

Unlike the single chain Heisenberg model, the undoped two chain ladder
does not have gapless spin $1/2$ excitations. Spin-charge separation does
not occur in the two chain ladder, and an extra up or down spin is
bound to a single hole, forming a quasiparticle. It is not possible to
specify a precise location for the extra spin, since every spin
fluctuates between up and down, but one can get some indication
of its whereabouts by measuring $\langle \psi_h | S^z_i | \psi_h \rangle$. 
In Fig. 2(b) we show this quantity for the same $|\psi_{h}\rangle$
shown in Fig. 2(a), in which the extra spin points up. 
Clearly the extra spin is localized close to
the hole, spending most of the time on the same rung as the hole.
Short-range antiferromagnetic correlations cause
$\langle S^z_i\rangle$ to be negative for some of the nearby sites.

In Fig. 2(c)-(d) we show similar results for a {\it static }
hole\cite{sandvik}.
In this case we remove one site, find the ground state, and measure
its properties. In Fig. 2(c) we see that there is no dimerization of the
bonds above the vacancy, and no antiferromagnetic correlations
between next-nearest-neighbor sites. In Fig. 2(d), we see that the
extra spin is still mostly on the same rung as the static hole, but
there is substantially more antiferromagnetic polarization caused by
the extra spin. For a dynamic hole, this polarization is mostly
absent because it reduces the overlap between adjacent hole
configurations, since it is tied to the hole location.

The addition of a static hole increases the total exchange energy
of the system, including the $-{1\over4}J n_{i} n_{j} $ term,
by $0.995t = 1.99 J$. The frustrating effects of a
dynamic hole further increase the total exchange energy by $0.26t$,
but the kinetic energy associated with the dynamic hole is $-2.37t$.

Next we consider the two chain system with two holes, which bind
to form a pair. In Fig. 3(a) we show the expectation value of the
kinetic energy on each bond, when the location of only one of the
holes has been specified with the projection operator $P_h(h)$.
This provides information not only on where the other hole is,
but between which sites it hops the most. We see that the other
hole spends most of its time on the opposite chain, close
to the first hole. What fraction of the time the other hole
spends on each of the sites is obtained from 
$\langle \psi | P_h(h)|\psi\rangle$ with both locations specified.
We find if the first hole is at (16,1), the probability for finding
the other on (15,2), (16,2), or (17,2), is about 0.15, for
(14,2) or (18,2), about 0.075, and for (15,1) or (17,1), about
0.055. The second hole spends more total time on the {\it two}  sites a
distance $\sqrt{2}$ away from the first hole than on the {\it three}  
nearest-neighbors sites. 
The probability is over 0.99 that the other hole is within
a distance of 6 of the first hole. 

Hole-hole density correlation functions have been calculated 
using Lanczos methods for
two holes on clusters ranging from $4\times4$\cite{pre,hp} up to
$\sqrt{26}\times\sqrt{26}$\cite{poilblanc}, and using Green's
function Monte Carlo techniques on an $8\times8$ cluster\cite{bon}.
It has been estimated that $J/t$ must be larger than $0.27$ for pair
binding to occur\cite{bon}. Near-neighbor and next-nearest-neighbor
diagonal hole-hole correlations are dominant for $J/t > 0.4$.
Based on the $\sqrt{26}\times\sqrt{26}$ results\cite{poilblanc},
for $J/t = 0.5$ the holes are about 20\% more likely to be found
across a diagonal than on near-neighbor sites. For $J/t$ greater
than about $1.0$, the near-neighbor correlation exceeds the diagonal
one.

In Fig. 3(b)-(d) we show the
bond strengths when the dynamic holes are in three possible
configurations. The exchange energy of $|\psi_h\rangle$, compared to
the system without holes, is 1.45$t$ for (b), 1.71$t$ for (c), and
1.92$t$ for (d).
Despite these energies, the system
spends as much time in configuration (d) as in (b), and much more time
in either of these than in (c). Configuration (d) is favored, despite
its high exchange energy, because it connects with more hole
configurations $h'$ than does (b) or (c), giving it more weight in the
kinetic part of the energy. There are six configurations $h'$ connected
to (d), but only four to (b) or (c) (counting hops of either hole).

Perhaps the most remarkable aspect of Fig. 3 is the very strong
next-nearest-neighbor singlet bond crossing the holes in Fig. 3(d). 
For four of the six hops available to (d), this bond becomes a 
nearest-neighbor bond, and in each of those neighboring configurations
the bond is quite strong. Therefore, the kinetic term strongly 
favors a singlet bond connecting these sites. 

On the $32\times 2$ system,
the kinetic energy of a pair of holes is -4.57, compared with
-4.74 for two separate holes. The increase in exchange energy
caused by a pair of holes is 2.06, compared with 2.51 for two
separate holes. Thus the slight increase in kinetic energy from binding
a pair of holes is more than made up for by a substantial 
decrease in exchange energy. The pair binding energy, defined as 
\begin{equation}
E_b = 2 E(1) - E(2) - E(0),
\end{equation}
where $E(m)$ is the ground states energy with $m$ holes, is
$E_b = 0.28$.

It is useful to define the {\it frustration energy}  associated
with a particular hole configuration $h$ as
\begin{equation}
E_f(h) = \langle \psi_h | {\cal H}_S | \psi_h \rangle - 
 \langle \psi_{h,static} | {\cal H}_S |
 \psi_{h,static} \rangle,
\end{equation}
where $| \psi_{h,static} \rangle$ is the ground state in the
static hole configuration $h$. The frustration energy of two
separate holes on the $32\times 2$ is $0.52t$. For two holes
in configurations (b), (c), and (d), the frustration energies are
0.077$t$, 0.249$t$, and 0.066$t$, respectively, much less than
for separate holes.
The frustration energy associated with the strong diagonal
frustrating bond in (d) is rather small.
This reflects the
fact that a free $S=1/2$ forms on the end of a two chain Heisenberg
ladder with one extra site on one chain \cite{rvbprl}, and this
extra spin can be used to form the diagonal singlet bond.

\subsection{Three chains}
We next consider a three chain ladder system with a single hole.
In Fig. 4(a) we show the kinetic energy on each link in a
$16\times3$ system, for
sites near the center of the system. It is clear that the hole
resides mostly on the outer legs, and that when it does hop onto a
center-leg site, it is most likely to then hop to an outer leg.
In Fig. 4(b) we show the exchange energy on each link near a
mobile hole in the state $|\psi_{h}\rangle$ with $h$ on an outer
leg near the center of the system. The expected next-nearest
neighbor antiferromagnetic bond has formed across the hole.
The dimerization is quite different than in the two-chain case:
it forms in the vertical direction, where it is both more effective
at accomodating the frustration and less costly in energy. The
dimerized bonds form a structure resembling a short two-leg ladder.
In Fig. 4(c) we show similar results for the hole on the adjacent
center-leg site. A particularly strong singlet bond forms across
the hole, reflecting a strong tendency to hop vertically. 
Fig. 4(d) shows $\langle S^z \rangle$ on sites about the
hole of Fig. 4(b). The pattern strongly resembles that of a single
chain. Instead of being localized near the hole, the extra $S=1/2$
is distributed about the system, indicating spin-charge separation.
The spins form a one-dimensional shifted-N\'eel configuration. 
On the same leg as the hole, the other two sites are bound tightly
into a singlet, and $\langle S^z \rangle$ is very small. Fig. 4(e) shows
similar results for the hole on the center leg.
The frustration energy of a single hole on an outer leg on
three chains is 0.19$t$, and on the center leg, 0.68$t$.

Two holes on a long three-chain ladder with $J/t=0.5$ are not bound.
The density of holes has two widely spaced broad peaks.
Fig. 4(f) shows the kinetic energy of one hole when the other
hole is projected onto a site at one of the peaks in the density.
Direct measurement of the hole-hole correlation function shows that
the hole is found exactly where the kinetic energy is large.

Why are two holes bound on two chains and not on three chains?
This is what one would expect based on arguments using an 
RVB variational ansatz for the background spin system\cite{rvbprl}.
However, those arguments are based on a static treatment of holes,
and as we have seen here, for $J/t=0.5$ the holes cannot be treated
statically. The RVB ansatz, as well as various analytical 
approaches\cite{ricetwo,sierra}, predicts the existence of free spinon
excitations on ladders with odd numbers of legs, and this is important.
In the three chain system a hole can separate
into a hole and a free, zero-energy spinon, 
which one would expect to have lower
energy. It is interesting to compare Fig. 4(b) and Fig. 2(a): 
on three chains, a low energy local spin 
structure can form involving vertical dimerization which allows
easy hopping. On two chains, vertical dimerization is not possible,
and the bonds above the hole also carry the extra $S=1/2$, reducing
their strength. Direct comparison of the energies supports this
picture: on a $16\times3$ system, adding one hole (and spinon) increases
the exchange energy by 1.19$t$, and decreases the kinetic energy
by 2.54$t$. On a $16\times2$ system, the exchange energy is increased
by 1.26$t$ and the kinetic energy decreases by 2.36$t$. By this measure
a pair of separated holes is lower in energy on three chains by 0.50$t$. 
The pair binding energy on the two chain system is 0.28$t$, less than
the difference in hole energy between two and three chains.

\subsection{Four chains}
We now consider a four chain ladder system with a single hole.
Unlike the three chain case, the probability of finding the hole
on the center two chains is about the same as finding it on an
outer chain. This is despite the fact that only three bonds are 
broken when the hole is on an outer chain, versus four for an inner
chain. In Fig. 5(a,b) we show the bond strengths about a dynamic hole
on an outer chain and on an inner chain.
Next-nearest-neighbor antiferromagnetic bonds have formed across the hole.
Dimerization is also present. The frustration energy of the hole
locations shown in Fig. 5(a) and (b) are 0.26$t$ and 0.42$t$, respectively.
The additonal spins surrounding the hole, compared to two or three
chains,  tend to reduce both
horizontal and vertical dimerization.  The precise pattern of frustrating
bonds and dimerization is somewhat complicated. One could imagine
putting in a static vacancy and including next-nearest-neighbor
interactions $J'$ about the vacancy to approximate the effect of
hole motion. This approach neglects the ability of the hole to
hop preferentially between some pairs of sites in order to adapt
to the frustration. This effect is visible in Fig. 5(b), where the
hole prefers to hop vertically rather than horizontally, and the
vertical frustrating bond is stronger. 

In Fig. 5(c,d) we show the same results, but with the undoped spin
background subtracted off. This indicates more clearly the
distortion caused by the hole. Notice that all the bonds
immediately surrounding the hole are weaker. In Fig. 5(e,f) we show
the same results for static holes (vacancies). The distortion of the spin
background for static holes is much smaller, and for the bonds  
immediately surrounding the hole, opposite in sign. The static
hole induces no dimerization or frustration.

We now consider two holes on a four chain ladder.
In Fig. 6 we show the expectation value of the
kinetic energy on each bond, when the location of only one of the
holes has been specified with the projection operator $P_h(h)$.
It is clear that the two holes are bound.
However, the precise pattern of hopping initially seems rather strange.
The patterns primarily reflect the fact that an {\it undoped }
two-leg ladder configuration of spins is a low energy configuration.
One can compare undoped ladders with even numbers of legs, which have
a spin gap, and odd numbers of legs, which are gapless. 
It is natural to expect that the gap comes
about both by a rise in the spin excitation energy {\it and} 
a lowering in the ``vacuum'' ground-state energy. Thus we expect
a two chain undoped ladder, which has a very large gap of about 
$0.5 J$, to be an especially low energy system in some sense. Hence
the two holes in Fig. 6 prefer to lie on either the top two legs or
the bottom two legs, or the top and the bottom legs, but not on the
first and third, second and third, or second and fourth. If a hole
is on the ``wrong leg'', it especially doesn't like to hop
horizontally.

A four chain undoped system will also
have low energy, since it too has a spin gap.
Tsunetsugu, et. al.\cite{tsunetsugu} have argued that this
can lead to striped phases in which one dimensional lines of
holes divide ladders with even numbers of legs. Our results clearly
indicate that both single holes and pairs of holes often
arrange their motion so that undoped two-leg ladder-like
arrangements of spins can form. The tendency toward formation of
four-leg ladder structures is weaker. 

The three chain results can also be interpreted in terms of ladder
formation, in that the holes predominantly sit on the outer legs,
with the other two legs near each hole forming an undoped two chain system.
An important difference
between the structures we see and those suggested by Tsunetsugu, et.
al., is in the density of holes or pairs of holes adjacent to
the ladders: we see quite low densities, while they suggested a
line of holes with density near unity.

In Fig. 7 we show the bond strengths about the two most probable
hole configurations, which are almost equally probable.
In (a) we see the strong next-nearest
neighbor diagonal singlet bond crossing the holes. Horizontal 
hopping transforms this diagonal bond into 
vertical bonds which sit on each side of the pair, as seen in (b).

The kinetic energy of a pair of holes is -5.16$t$, compared with
-5.25$t$ for two separate holes. The increase in exchange energy
caused by a pair of holes is 2.47$t$, compared with 2.78$t$ for two
separate holes. As in the two-chain case, the slight increase in 
kinetic energy from binding a pair of holes is more than made up for
by the decrease in exchange energy. The pair binding energy is 
$E_b = 0.21t$. This pair binding energy is smaller by 25\% than
the two chain value. In contrast, the spin gap for the undoped four
chain system is smaller by over 60\% compared to two chains.
The frustration energy corresponding to Fig. 7(a) is 0.20$t$. For
Fig. 7(b), it is 0.14$t$. The frustration energy of two separate
holes would be 0.51$t$ if they were both on outer legs, and 
0.68$t$ if one was on an outer leg and one on an inner leg.

\subsection{Five chains}
We now consider a five chain ladder system with a single hole.
Recall that a single hole on a three chain ladder spent most of
the time on an outer chain. Since the undoped three and five chain
systems have similar, gapless ground states, one might expect 
a hole on the five chain system to spend more time on an outer
chain than in the center. However, this is not the case. In
Fig. 8(a) we show the kinetic energy for a single hole on
an $8\times5$ cluster. The hole spends most of the time on the center
chain. By moving on the center chain, the system is divided into
two undoped two-chain ladder systems above and below the hole.
As Fig. 8(b) shows, this configuration allows the vertical
dimerization found in the three chain system to form both above
and below the hole, allowing a strong frustrating bond to form
horizontally across the hole. The hole tends not to hop all the
way to the ends of the system so that vertical two-chain structures
can form there. The frustration energy at site (4,3) is $E_f = 0.26t$.
In Fig. 8(c), the corresponding
spin configuration is shown. The same shifted N\'eel pattern found
in three chains is again seen, with the spins on an entire five-site rung
shifting with the motion of the hole.

Two holes in this system repel. The spin configuration around a
single hole is highly favorable, as in the three chain case. The
``core'' of a bound pair of holes is a $2\times2$ plaquette. If 
a pair were to form, it would divide the system into a two chain
ladder and a single chain, and the single chain would have high energy.
In fact, the separate holes form the structure shown in Fig. 9(a),
where the system is divided into ladders both horizontally and
vertically! In Fig. 9(b), we show the kinetic energy of one of
the holes when the other is projected onto a site. The holes
clearly are unbound. The system dimerizes vertically above and
below each hole, and horizontally to the left and right of
each hole.

The vertical hopping patterns are highly dependent on the length of
the system. An $8\times5$ system with two holes allows convenient 
division of the system into width-two pieces in both directions.
In Fig. 9(c,d), we show the results similar to those shown in Fig.
9(a,b) but for a $10\times5$ system. In this case, some of the
vertical pieces must be of width greater than two. In Fig. 9(d),
we see that when the first hole is on site (3,3), the motion
of the second holes divides the right part of the system into
either two horizontal two-chain ladders or into a vertical two-chain
and a vertical four-chain ladder.

\subsection{A $8\times6$ cluster}
We now consider an $8\times6$ cluster with a single hole.
In Fig. 10(a) we show the kinetic energy per bond for the hole.
A single hole is likely to be found in the central sites of
the cluster, allowing a two leg ladder to run along the entire edge of
the system. Some slight asymmetry is visible in the figure; this
is a result of a slight numerical inaccuracy in the DMRG
calculation. We kept 600 states per block in this calculation; 
despite this many states, the truncation error (also refered to as
the discarded weight) was relatively high: 
about $6\times10^{-5}$. This level of accuracy was, however, 
sufficient to determine the general structure of the hole.
In Fig. 10(b) we show the bond strengths about the hole projected
onto a central site. 
Next-nearest-neighbor antiferromagnetic bonds have formed across the
hole, but they are somewhat weaker than in the narrower systems.
This reflects a decreased ability to hop in this system, which is
dominated more by the exchange energy.
Dimerization is also present, particularly beneath the hole, but it
is also weaker than in the narrower systems. In Fig. 10(c), we show
the difference in bond strength between the one hole system and the
undoped system. The distortion caused by the hole is fairly
substantial over a $5\times5$ region. All the bonds
immediately surrounding the hole are weaker. In Fig. 10(d) we show
the same results for a static hole. The distortion of the spin
background is much smaller than for a dynamic hole 
(note the decreased scale), and for the bonds  
immediately surrounding the hole, opposite in sign. 
The frustration energy for this hole location is $E_f=0.29$.

In Fig. 11 we show results for two holes on an $8\times6$ cluster.
Again we kept 600 states per block, but the truncation error was 
higher than in the single-hole calculation: $2\times10^{-4}$. This
was still sufficient to determine the structure of the pair with
reasonable accuracy and to determine the pair binding energy.
In Fig. 11(a) we show the expectation value of the
kinetic energy of a hole, when the other hole has been projected
onto a central site. The two holes are clearly bound. 
The hole is somewhat more likely to be to the left of the projected
hole than one might expect; however, this configuration breaks
the system vertically into two-chain and four-chain undoped ladders,
rather than two three-chain undoped ladders. 

In Fig. 11(b-d) we show the bond strengths surrounding
several likely configurations of the pair. The frustrating diagonal
singlet crossing the pair is clearly present in Fig. 11(b): this 
is the clearest ``signature'' of a bound pair of holes, and is
present in all the systems in which we have found pair binding.
In addition, additional frustrating bonds crossing the holes are
present in both directions. Vertical dimerization is present above
and below the holes, where it is expected, and to the left and
right, where we might have expected horizontal dimerization.
Even on a system of width 6, the boundaries are still substantially
affecting the spin structure surrounding the pair, and it is not clear which
type of dimerization would appear in a large system. The most 
probable configuration of the pair is not shown: (3,4)-(4,3), with
probability 0.018. Configuration (b), (c), and (d) have
probabilities of 0.014, 0.005, and 0.017, respectively.
The frustration energies $E_f(h)$ for configurations (b), (c), and
(d) are 0.32$t$, 0.54$t$, and 0.20$t$, respectively. The frustration 
energy of two separate holes is 0.58$t$.

In Fig. 11(e,f) we show the difference in bond strengths of the
two-hole system and the undoped system. Substantial distortion
of the spin structure occurs over a $6\times6$ region.

The kinetic energy of a pair of holes on an $8\times6$ cluster
is -5.36$t$. Twice the kinetic energy of a single hole is -5.38$t$. 
The increase in exchange energy
caused by a pair of holes is 2.71$t$, compared with 2.96$t$ for two
separate holes. The increase in 
kinetic energy from binding a pair of holes is very tiny, and is
more than made up for by the decrease in exchange energy. 
The pair binding energy is 
$E_b = 0.24(2)t$. This pair binding energy is slightly bigger than
on a $16\times4$ lattice.

\subsection{$8\times7$ and $10\times7$ clusters}

We have performed a few DMRG studies of width 7 systems. We studied
a $10\times7$ cluster with two holes, keeping 800 states per block,
with a total of 10 sweeps through the lattice.
The truncation error was fairly large, $2\times10^{-4}$, but it was
clear that the two holes were bound, and tended to stay near the
center of the cluster. In general, the results were similar
to the $8\times6$ cluster. We also studied a $8\times7$ cluster with
a staggered magnetic field $H=0.15$ applied to the edge sites. 
The idea was to simulate the N\'eel spin background of an infinite 
undoped lattice.  The field strength was chosen to represent a mean
field coupling to surrounding sites, each with an average
magnetization of $\langle S^z_i\rangle=0.3$. 
Again, two holes were bound, with a pair
binding energy of about $0.15$. 

\subsection{A $2\times2$ cluster}
The bound pair of holes which have been found in a number of these
clusters are characterized by a $2\times2$ core region over which
the dominant hole-hole correlations occur. In order to better
understand this core we consider the $2\times2$ lattice shown in 
Fig. 12.

Introducing the singlet valence bond operator between sites $i$ and
$j$
\begin{equation}
\Delta_{ij}^\dagger  = \frac{1}{\sqrt{2}} (
c^\dagger_{i,\uparrow} c^\dagger_{j,\downarrow} +
c^\dagger_{j,\uparrow} c^\dagger_{i,\downarrow}) ,
\end{equation}
the ground state of the undoped half-filled system can be written as 
\begin{equation}
|\psi\rangle_0 = N_0 [ \Delta_{14}^\dagger \Delta_{23}^\dagger -
\Delta_{12}^\dagger \Delta_{34}^\dagger ] |0\rangle ,
\label{undoped}
\end{equation}
with $|0\rangle$ the vacuum.
The ground state of the two-hole system is
\begin{equation}
|\psi\rangle_2 = N_2 [ 
a(\Delta_{12}^\dagger + \Delta_{23}^\dagger +
\Delta_{34}^\dagger + \Delta_{14}^\dagger) +
b(\Delta_{13}^\dagger + \Delta_{24}^\dagger)
] |0\rangle ,
\label{doped}
\end{equation}
with $a=1$ and $b=[2+(J/4t)^2]^{1/2} - J/4t$.
In the doped, two-hole state $|\psi\rangle_2$, the ratio of the
edge singlet (e.g. 1-2) to diagonal singlet (e.g. 1-3) amplitude is
\begin{equation}
\frac{a}{b}=\frac{1}{[2+(J/4t)^2]^{1/2} - J/4t}.
\end{equation}
For $J/t=2$, this ratio is unity. For $J/t<2$, the diagonal 
amplitude is larger than the edge amplitude. This is reflected in
the $t$-$J$ results previously discussed, where for $J/t=0.5$
the hole-hole 
correlations were found to be larger for next-nearest-neighbor
diagonal sites than for nearest-neighbor sites.

The ground state, Eq. (\ref{undoped}), of the undoped 
$2\times2$ system transforms as $d_{x^2-y^2}$, while the two-hole
state, Eq. (\ref{doped}), transforms as an $s$-wave. Thus the hole-pair
creation operator that connects $|\psi\rangle_0$ to $|\psi\rangle_2$ 
must tranform as $d_{x^2-y^2}$\cite{moreo,trugman}. A simple 
nearest-neighbor operator of this form is 
\begin{equation}
\Delta  = \Delta_{14} - \Delta_{12} + \Delta_{23} - \Delta_{34}.
\label{nnd}
\end{equation}
Applying this to the undoped ground state $|\psi\rangle_0$ given by
Eq. (\ref{undoped}), one finds that 
\begin{equation}
\Delta |\psi\rangle_0  = -2N_0[
\Delta_{12}^\dagger + \Delta_{23}^\dagger +
\Delta_{34}^\dagger + \Delta_{14}^\dagger]|0\rangle  .
\end{equation}
This clearly has a nonzero overlap with with the two-hole ground
state $|\psi\rangle_2$, but it does not contain the diagonal singlet
parts. However, if we were to ``time-evolve'' this state towards
the two-hole ground state by applying $e^{-{\cal H}\tau}$ we would have
for short imaginary times $\tau$
\begin{equation}
e^{-{\cal H}\tau} \Delta |\psi\rangle_0 \approx 
(1-{\cal H}\tau) \Delta |\psi\rangle_0 ,
\label{expH}
\end{equation}
and the hopping kinetic energy in ${\cal H}_K$ generates the diagonal singlet
terms 
\begin{equation}
{\cal H}_K \Delta |\psi\rangle_0 \sim 
t ( \Delta_{13}^\dagger + \Delta_{24}^\dagger)|0\rangle .
\end{equation}
An improved hole pair creation operator would include, in addition
to $\Delta$, terms of the form ${\cal H}_K\Delta$.

A $d_{x^2-y^2}$ hole pair creation operator, generalized to include
holes on next-nearest-neighbor diagonal sites, has been discussed
by Poilblanc\cite{poilblanc}. 
One can expand
a generalized hole-pair creation operator in terms of operators
which create a pair of holes on sites separated by a distance $R$. 
For our $2\times2$ cluster this involves 
\begin{equation}
\Delta_{d_{x^2-y^2}} = \sum_R \Delta_R,
\label{sumr}
\end{equation}
with $R=1$ and $R=\sqrt{2}$.
The nearest-neighbor operator $\Delta_1$ is just the operator given
in Eq. (\ref{nnd}). As discussed by Poilblanc, a 
next-nearest-neighbor term possessing $d_{x^2-y^2}$ symmetry is
\begin{equation}
\Delta_{\sqrt{2}} = (\vec S_1 - \vec S_3) \cdot \vec T_{24}
- (\vec S_2 - \vec S_4) \cdot \vec T_{31},
\end{equation}
with
\begin{equation}
\vec S_1 \cdot \vec T_{24}
= \frac{1}{2} 
(c_{1\uparrow}^\dagger c_{1\uparrow} - c_{1\downarrow}^\dagger c_{1\downarrow})
(c_{2\uparrow} c_{4\downarrow} - c_{4\uparrow} c_{2\downarrow})
+ c_{1\uparrow}^\dagger c_{1\downarrow} c_{2\uparrow}c_{4\uparrow}
+ c_{1\downarrow}^\dagger c_{1\uparrow}
c_{2\downarrow}c_{4\downarrow} .
\label{pd}
\end{equation}
Note that since $\vec T_{24}=-\vec T_{42}$, Eq. (\ref{pd})
has $d_{x^2-y^2}$ symmetry.
Acting on the undoped ground state, $\Delta_{\sqrt{2}}$ generates
the diagonal singlets 
\begin{equation}
\Delta_{\sqrt{2}}  |\psi\rangle_0 \sim 
( \Delta_{13}^\dagger + \Delta_{24}^\dagger)|0\rangle .
\end{equation}
The action of this operator on $|\psi\rangle_0$ is the same as
${\cal H}_K\Delta$.

Based on this, we believe that the bound hole pairs observed in
various clusters should be thought of as $d_{x^2-y^2}$ pairs.
The diagonal-singlet bond as well as the nearest-neighbor singlet bonds
reflect the two-hole structure of Eq. (\ref{doped}). In the larger
clusters the pair structure is more extended, corresponding to
further operations of $(1-{\cal H}\tau)^N$ on $ \Delta
|\psi\rangle_0$, Eq. (\ref{expH}), 
or longer-range operators $\Delta_R$ in Eq. (\ref{sumr}). The pair
structure on larger systems includes both larger separation of
the holes and alterations of the spin background near the pair.

\section{Discussion}
In considering one and two hole ground states of a wide variety of
clusters, we have found a remarkable sensitivity to the shape of
the cluster. Underlying the variety of results, however, are a few
basic low-energy structures. The nature of the ground state of any
particular system is based on which arrangement of these basic
structures is lowest in energy. 

The most important structure is a bound pair of holes. This
structure allows the pair to hop rather freely in order to decrease
the kinetic energy, without disrupting the spin background more than
necessary. The bound pair is characterized by a $2\times2$ ``core''
region discussed above. Surrounding the core and extending several
lattice spacings further is a region in which the
spin structure is strongly perturbed. Within the core, for
the case $J/t=0.5$ which we have studied, the pair of
holes is more likely to be at next-nearest-neighbor diagonal sites
than nearest-neighbor sites, in order to maximize the hopping
overlap with other hole configurations. When the two holes are diagonally
situated, a strong singlet bond is present across the other two
sites of the core. This singlet becomes a strong nearest-neighbor
singlet bond after one of the holes hops next to the other. The
singlet forms in order to maximize the hopping overlap with
these other hole configurations. In order to respond to this
frustrating bond, and to other weaker frustrating bonds across each
of the holes, the surrounding spins dimerize, reducing
the spin-spin correlations around the pair. The effect of this
dimerization is to induce a ``spin-liquid'' region surrounding the pair.

Frustrating next-nearest-neighbor bonds forming across holes are
a universal feature in all of the clusters we have studied. 
These bonds are necessary for
hole motion.  Holes bind in pairs in order to share their
frustration. This mechanism for pairing is quite different from
simple ``broken-bond'' counting, which predicts nearest-neighbor
pairing for {\it static}  holes: for two static holes, a nearest
neighbor configuration eliminates seven bonds, while anything else
eliminates eight. For physical values of $J/t$, such as $J/t=0.5$,
the ``broken-bond'' effect enhances pair-binding somewhat, but is 
not dominant. Consider once again the $8\times6$ cluster, with 
two holes.   Results for the hole-hole correlation
function indicate that the pair resides on nearest-neighbor sites
only 22\% of the time. Even if a broken bond results in an extra exchange
energy of $J=0.5$, the effect on pair binding is only $0.22J = 0.11t$, while
the actual pair binding energy is $0.24(2)t$. A more accurate
estimate of the effect of broken bonds comes from considering two
{\it static} holes on an $8\times6$ cluster: the difference in
energy between nearest-neighbor static holes and widely separated
holes is $0.62J$, rather than $J$, suggesting that the broken-bond
energy for dynamic holes is about $0.07t$. In contrast, the
frustration energy for two bound holes in the most probable hole
configurations ranges from 0.25$t$-0.40$t$ {\it less than}  the 
frustration energy of two separate dynamic holes.

The next most important structure is a nearly undoped two-leg ladder
region. The large spin gap of a two-leg ladder coincides with
a low energy spin-liquid ground state. The two-leg ladder is
dimerized, in that the rung bonds are stronger than the leg bonds.
This makes the ladder especially suited for a hole or pair of
holes to move beside it. 

Usually when ladder structures form,
they are bounded by regions with pairs. In the special case of a five-leg
ladder, pairs of holes are too wide, and the system instead 
has unpaired holes moving in one dimension, breaking the system
into two-leg ladders.  This is the last important structure:
a one-dimensional line of unpaired holes.
This structure is low enough in energy
that it can appear in order to allow the formation of one or two
undoped two-leg ladders, specifically in the three and five chain
systems.

The energy difference between these structures is sufficiently small
that modest external perturbations can lead to the trapping of
holes or the formation of static even-leg ladders. 
Even in the absence of external perturbations, we would expect that
a dilute concentration of holes will give rise to fluctuating
extended structures in the medium. 
Although we have not presented results here for more than
two holes, we have such results for three, four, and five chains and will
present them elsewhere. 
Based on our results at finite doping, we believe that the
tendency to form two-leg ladders persists into the finite, but low doping
regime, while at moderate doping, ladders diminish in importance and 
pairs of holes dominate.
The even-leg ladders that are present in the dilute system could
give rise to the pseudo-gap observed in the underdoped cuprates.
It has been suggested that holes doped into the $t$-$J$ model will
phase separate\cite{emerykl}. However,
in our studies with more holes, we find no evidence for phase
separation at $J/t=0.5$, which is consistent with previous $t$-$J$ studies on
clusters up to $\sqrt{26}\times\sqrt{26}$, in which 
phase separation was found only at larger $J/t$
values\cite{poilblanc}. The ladder structures we find are not the
result of a competition between phase separation and a long-range
Coulomb interation\cite{emeryk}, but arise directly from the
short-range interactions in the $t$-$J$ model.

\section*{Acknowledgements}

We acknowledge support from the from the NSF under Grant Nos.
DMR-9509945 and DMR95--27304.
Some of the calculations were performed at the
San Diego Supercomputer Center.

\begin{figure}
\caption{Spin structure near a single hole (the gray circle) on a 1D
$t$-$J$ lattice.
(a) N\'eel spin configuration, shifted by one spacing to the right of
the hole.
(b,c) Valence bond configurations with a hole.
(d,e) Results of a DMRG calculation for the ground state of
a 15 site $t$-$J$ system, with $J/t = 0.5$, and open boundary
conditions.  The thickness of the lines
is proportional to the bond strengths,
$\langle \psi | \vec S_i \cdot \vec S_j P_h(h)|\psi \rangle / 
\langle \psi | P_h(h)|\psi \rangle$, 
according to the scale shown.
In (d), $h=7$, and in (e), $h=8$.  
(f) Results of a DMRG calculation for the ground state of a 16 site
system, with $J/t = 0.5$, and open boundary conditions. 
The length of the arrow is proportional to $\langle S^z
P_h(h)\rangle/\langle \psi | P_h(h)|\psi \rangle$.
}
\label{one}
\end{figure}

\begin{figure}
\caption{A single hole on a two-chain ladder. Gray circles are
dynamic holes, and black circles are static vacancies.
Pictured is the central region of a $32\times2$ lattice, with open
boundary conditions. All results are for $J/t=0.5$.
(a) The bond strengths $\langle \vec S_i \cdot \vec S_j \rangle$
about a dynamic hole, as in Fig. 1(d)-(e). All nearest-neighbor
bonds are shown. In addition, if two sites are both adjacent to
the hole, and if the bond is antiferromagnetic,
$\langle \vec S_i \cdot \vec S_j \rangle < 0$, it is also shown.
(b) $\langle S^z \rangle$, as in Fig. 1(f). 
(c) The bond strengths $\langle \vec S_i \cdot \vec S_j \rangle$
about a static vacancy.
(d) $\langle S^z \rangle$ about a static vacancy.
}
\label{two}
\end{figure}

\begin{figure}
\caption{Two dynamic holes (gray circles) on a two-chain ladder. 
Pictured is the central region of a $32\times2$ lattice, with open
boundary conditions.
(a) The hopping energy 
$- t \displaystyle \sum_s \langle c^\dagger_{i,s}c_{j,s} + 
c^\dagger_{j,s}c_{i,s}\rangle $ for each link when one hole is projected onto
a particular site. The hopping energy shown is associated with
the hole which has not been projected onto a particular site.
The thickness of the lines is proportional to energy, according
to the scale shown.
(b-d) The bond strengths $\langle \vec S_i \cdot \vec S_j \rangle$
after both holes have been projected.
Next-nearest-neighbor bonds are shown if $i$ and $j$ are both adjacent to
the same hole, and if the bond is antiferromagnetic.
}
\label{three}
\end{figure}

\begin{figure}
\caption{Dynamic holes on a three-chain ladder, plotted
similarly to Figs. 2 and 3. Parts (a)-(e) are for a single
hole, and (f) is for two holes.
Pictured is the central region of a $16\times3$ lattice, with open
boundary conditions.
(a) The hopping energy for each link.
(b,c) The bond strengths $\langle \vec S_i \cdot \vec S_j
\rangle$.
(d,e) $\langle S^z \rangle$ for each site about the hole.
(f) The hopping energy in a two-hole system when one hole
has been projected onto one of its most probable locations.
}
\label{four}
\end{figure}

\begin{figure}
\caption{A single dynamic hole on a four-chain ladder.
All calculations are on a $16\times4$ lattice, with open
boundary conditions. Only the central region is shown.
(a,b) The bond strengths $\langle \vec S_i \cdot \vec S_j
\rangle$ for two different hole locations.
(c,d) The difference in bond strengths $\langle \vec S_i \cdot \vec S_j
\rangle$ between the system with the dynamic hole [as in (a) and
(b)] and the same bond on the equivalent {\it undoped } system. 
Solid lines indicate stronger bonds, and dashed indicate weaker. 
Next-nearest-neighbor bonds are not shown.
(e,f) The difference in bond strengths between the system with a
{\it static} hole or vacancy (black circle) and the equivalent undoped system.
}
\label{five}
\end{figure}

\begin{figure}
\caption{Two dynamic holes on a four-chain ladder.
The figure shows the hopping energy of one dynamic hole when the other has
been projected onto a particular site.
All calculations are on a $16\times4$ lattice, with open
boundary conditions. Only the central region is shown.
}
\label{six}
\end{figure}

\begin{figure}
\caption{Exchange energy for a $16\times4$ system, with open
boundary conditions, and two dynamic holes.
}
\label{seven}
\end{figure}

\begin{figure}
\caption{A single dynamic hole on a five-chain ladder.
All calculations are for an $8\times 5$ lattice, with open 
boundary conditions.
(a) The hopping energy of the hole.
(b) The bond strengths about the hole.
(c) $\langle S^z \rangle$ for each site about the hole.
}
\label{eight}
\end{figure}

\begin{figure}
\caption{Two dynamic holes on a five-chain ladder.
(a) The hopping energy for an $8\times 5$ cluster, with open
boundary conditions.
(b) The hopping energy for the system shown in (a) 
with one hole projected onto a site.
(c) The hopping energy for an $10\times 5$ cluster, with open
boundary conditions.
(d) The hopping energy for the system shown in (c)
with one hole projected onto a site.
}
\label{nine}
\end{figure}

\begin{figure}
\caption{A single dynamic hole on a $8 \times 6$ system, with
open boundary conditions.
(a) The hopping energy of the hole.
(b) The bond strengths about the hole.
(c) The difference in bond strengths between the system with the dynamic hole 
and the same bond on the equivalent undoped system.
Solid lines indicate stronger bonds, and dashed indicate weaker. 
(d) The difference in bond strengths between the system with a
{\it static} hole or vacancy (black circle) and the equivalent undoped system.
}
\label{ten}
\end{figure}

\begin{figure}
\caption{Two dynamic holes on a $8 \times 6$ system, with
open boundary conditions.
(a) The hopping energy for each link when one hole is projected onto
a particular site. 
(b-d) The bond strengths about the pair of holes.
(e,f) The difference in bond strengths between the system with the dynamic hole 
and the same bond on the equivalent undoped system.
}
\label{eleven}
\end{figure}

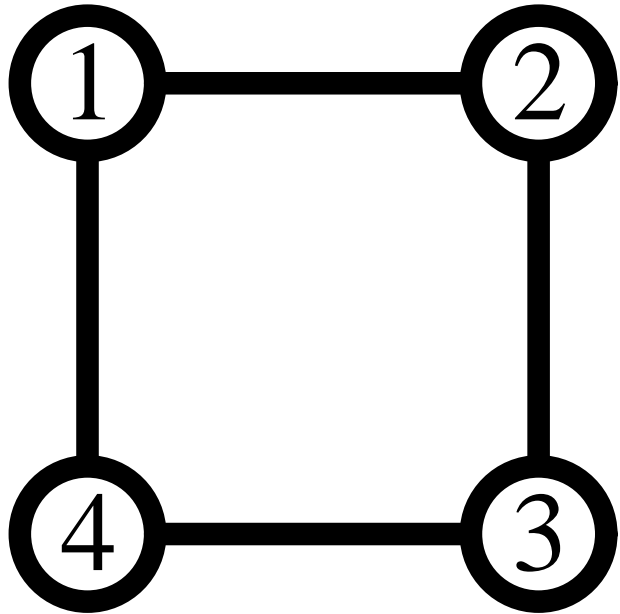
\begin{figure}
\caption{The $2 \times 2$ $t$-$J$ cluster. Edge nearest-neighbor 
singlets can form as well as diagonal (1-3,2-4)
next-nearest-neighbor singlets.
}
\label{twelve}
\end{figure}


\newpage

\begin{references}
\bibitem{infinited} W. Metzner and D. Vollhardt, \prl 
{\bf 62}, 324 (1989); A. Georges and G. Kotliar, 
\prb {\bf 45}, 6479 (1992); M. Jarrell, \prl {\bf 69}, 168 (1992);
A. Georges, G. Kotliar, W. Krauth, and M. J. 
Rozenberg, Rev. Mod. Phys. {\bf 68}, 13, (1996).

\bibitem{bosonization} D.C. Mathis and E.H. Lieb, J. Math. Phys.
{\bf 6}, 304 (1965); A. Luther and V.J. Emery, \prl {\bf 33}, 589
(1974); H. Frahm and V. Korepin, \prb {\bf 42}, 10553 (1990).

\bibitem{qmc} D.J. Scalapino, {\it Perspectives in Many-Particle
Physics},  R.A. Broglia, J.R. Schrieffer, and P.F. Bortignon, eds.
(North Holland, 1994) 95-125.

\bibitem{pre} J. Bonca, P. Prelovsek, and I. Sega,
\prb {\bf 39}, 7074 (1989).

\bibitem{hp} Y. Hasegawa and D. Poilblanc, \prb {\bf 40}, 9035
(1989).

\bibitem{dag} E. Dagotto, J. Riera, and A.P. Young,
\prb {\bf 42}, 2347 (1990).

\bibitem{prelov} P. Prelovsek and X. Zotos,
\prb {\bf 47}, 5984 (1993).

\bibitem{poilblanc} D. Poilblanc, \prb {\bf 49}, 1477 (1994).

\bibitem{sign} E.Y. Loh, et.al., \prb {\bf 41}, 9301 (1990).

\bibitem{dmrg} S.R. White, \prl {\bf 69}, 2863 (1992),
\prb {\bf 48}, 10345 (1993).

\bibitem{cavo} S.R. White, 1996 preprint, cond-mat/9604129.

\bibitem{zhang} F.C. Zhang, T.M. Rice, \prb {\bf 37}, 3759 (1988).

\bibitem{ogata} M. Ogata and P.W. Anderson,  \prl {\bf 70}, 3087 (1993).

\bibitem{bursill}  R.J. Bursill , G.A. Gehring, 
D.J.J. Farnell, J.B. Parkinson, Chen Zeng, T. Xiang, preprint, 
cond-mat/9511044.

\bibitem{chitra} R. Chitra, S. Pati, H.R. Krishnamurthy, D. Sen, and
S. Ramasesha, \prb {\bf 52}, 6581 (1995).

\bibitem{zigzag} S. R. White and I. Affleck, preprint,
cond-mat/9602126.

\bibitem{siggia} B.I. Shraiman and E.D. Siggia, \prl {\bf 60}, 740
(1988).

\bibitem{sandvik} A. Sandvik, D.J. Scalapino, and E. Dagotto have
studied the magnetic properties of static holes in
antiferromagnetic Heisenberg ladders and 2D clusters (unpublished).

\bibitem{bon} M. Boninsegni and E. Manousakis, 
\prb {\bf 47}, 11897 (1993).

\bibitem{rvbprl} S.R.\ White, R.M.\ Noack, and D.J.\ Scalapino,
\prl {\bf 73}, 886 (1994).

\bibitem{ricetwo} M Reigrotzki, H. Tsunetsugu, and T.M. Rice,
J. Phys. Cond. Matt. {\bf 6}, 9235 (1994).
(1988).

\bibitem{sierra} G. Sierra, preprint, cond-mat/951207.

\bibitem{tsunetsugu} H. Tsunetsugu, M. Troyer, and T.M. Rice,
\prb {\bf 51}, 16456 (1995).

\bibitem{moreo} A. Moreo and E. Dagotto,
\prb {\bf 41}, 9488 (1990).

\bibitem{trugman} D.J. Scalapino and S.A. Trugman, to appear in
Philosophical Mag.

\bibitem{emerykl} V.J. Emery, S.A. Kivelson, and H.Q. Lin,
\prl {\bf 64}, 475 (1990) and \prb {\bf 42}, 6523 (1990).

\bibitem{emeryk} V.J. Emery and S.A. Kivelson, Physica C {\bf
207}, 597 (1993).


\end{references}
\end{document}